\documentclass[aps,prl,preprint,nopacs,superscriptaddress]{revtex4}
\usepackage{graphicx}
\usepackage{verbatim}
\usepackage{amsmath,amsfonts,amssymb}
\usepackage{graphicx}

\def\3{2.8in}    
\def\2{2.5in}
\def\4{3.0in}

\def \beq {\begin{equation}}
\def \eeq {\end{equation}}
\begin{document}

\title{Spin reorientation transition on the surface of a topological insulator}
\author{Su-Yang Xu}\affiliation {Joseph Henry Laboratory, Department of Physics, Princeton University, Princeton, New Jersey 08544, USA}
\author{M. Neupane}\affiliation {Joseph Henry Laboratory, Department of Physics, Princeton University, Princeton, New Jersey 08544, USA}
\author{Chang Liu}\affiliation {Joseph Henry Laboratory, Department of Physics, Princeton University, Princeton, New Jersey 08544, USA}
\author{D. M. Zhang}\affiliation {Department of Physics, Penn State University, University Park, PA 16802}
\author{A. Richardella}\affiliation {Department of Physics, Penn State University, University Park, PA 16802}

\author{L. A. Wray}\affiliation {Joseph Henry Laboratory, Department of Physics, Princeton University, Princeton, New Jersey 08544, USA}\affiliation {Advanced Light Source, Lawrence Berkeley National Laboratory, Berkeley, California 94305, USA}

\author{N. Alidoust}\affiliation {Joseph Henry Laboratory, Department of Physics, Princeton University, Princeton, New Jersey 08544, USA}
\author{M. Leandersson}\affiliation {MAX-lab, P.O. Box 118, S-22100 Lund, Sweden}
\author{T. Balasubramanian}\affiliation {MAX-lab, P.O. Box 118, S-22100 Lund, Sweden}

\author{J. S\'anchez-Barriga}\affiliation {Helmholtz-Zentrum Berlin f\"ur Materialien und Energie, Elektronenspeicherring BESSY II, Albert-Einstein-Str. 15, D-12489 Berlin, Germany}
\author{O. Rader}\affiliation {Helmholtz-Zentrum Berlin f\"ur Materialien und Energie, Elektronenspeicherring BESSY II, Albert-Einstein-Str. 15, D-12489 Berlin, Germany}

\author{G. Landolt}\affiliation {Swiss Light Source, Paul Scherrer Institute, CH-5232, Villigen, Switzerland}\affiliation {Physik-Institute, Universitat Zurich-Irchel, CH-8057 Zurich, Switzerland}
\author{B. Slomski}\affiliation {Swiss Light Source, Paul Scherrer Institute, CH-5232, Villigen, Switzerland}\affiliation {Physik-Institute, Universitat Zurich-Irchel, CH-8057 Zurich, Switzerland}
\author{J. H. Dil}\affiliation {Swiss Light Source, Paul Scherrer Institute, CH-5232, Villigen, Switzerland}\affiliation {Physik-Institute, Universitat Zurich-Irchel, CH-8057 Zurich, Switzerland}

\author{T.-R. Chang}\affiliation {Department of Physics, National Tsing Hua University, Hsinchu 30013, Taiwan}
\author{H.-T. Jeng}\affiliation {Institute of Physics, Academia Sinica, Taipei 11529, Taiwan}\affiliation {Department of Physics, National Tsing Hua University, Hsinchu 30013, Taiwan}

\author{J. Osterwalder}\affiliation {Physik-Institute, Universitat Zurich-Irchel, CH-8057 Zurich, Switzerland}

\author{H. Lin}\affiliation {Department of Physics, Northeastern University, Boston, Massachusetts 02115, USA}
\author{A. Bansil}\affiliation {Department of Physics, Northeastern University, Boston, Massachusetts 02115, USA}
\author{N. Samarth}\affiliation {Department of Physics, Penn State University, University Park, PA 16802}
\author{M. Z. Hasan}\affiliation {Joseph Henry Laboratory, Department of Physics, Princeton University, Princeton, New Jersey 08544, USA}

\pacs{}

\date{\today}
\maketitle

\textbf{
The surface of topological insulators is proposed as a promising platform for spintronics and quantum information applications. In particular, when time-reversal symmetry is broken, topological surface states are expected to exhibit a wide range of exotic \textit{spin} phenomena \cite{Moore, Kane PRB, Qi PRB, Yu Science QAH, Qi Science Monopole, Galvanic effect} for potential implementation in electronics. Such devices need to be fabricated using nanoscale artificial thin films. It is of critical importance to study the spin behavior of artificial topological thin films associated with magnetic dopants, and with regards to quantum size effects related to surface-to-surface tunneling as well as experimentally isolate time-reversal breaking from non-intrinsic surface electronic gaps. Here we present observation of the first study of magnetically induced spin reorientation phenomena on the surface of a topological insulator. Our results reveal dramatic rearrangements of the spin configuration upon magnetic doping contrasted with chemically similar nonmagnetic doping as well as with quantum tunneling phenomena in ultra-thin films. While we observe that the spin rearrangement induced by quantum tunneling occurs in a time-reversal invariant fashion, we present critical and systematic observation of an out-of-plane spin texture evolution correlated with magnetic interactions, which breaks time-reversal symmetry, demonstrating $E(\vec{k}=0,\uparrow){\neq}E(\vec{k}=0,\downarrow)$ at a Kramers' point on the surface.}


Unlike in theory, magnetism in actual topological materials exhibit quite complex phenomenology  \cite{Hor PRB BiMnTe, Yayu WAL, Fe XMCD, Chen Science Fe, Andrew Nature physics Fe, STM Fe,  Ando QPT}. Experimentally, although some forms of magnetism in TI material have been reported by bulk doping \cite{Hor PRB BiMnTe, Yayu WAL} or surface deposition doping \cite{Fe XMCD}, the signature of TR symmetry breaking, spin reorientation of the new groundstate correlated with ferromagnetism, or the spin texture of the tunneling gap, is entirely lacking. 
The simplest physical scenario used in all theoretical proposals \cite{Qi PRB, Yu Science QAH, Qi Science Monopole, Galvanic effect} is that of an energy gap opened at the Kramers' point via ferromagnetism. If the gap is due to TR breaking effect, then the spin texture at the edge of the gap at the Kramers' point will be abruptly modified developing out-of-plane Fermi surface and higher binding energy spin polarization on the surface states, independent of Fermi surface warping \cite{Liang Fu Warping}. These out-of-plane spin polarization especially near the gap edge is the critical evidence for TR symmetry breaking long-sought for all potential applications, and has not been observed in any of the experiments so far. Indeed, a number of experiments have been performed to address the electronic states near the Dirac point of TIs with bulk magnetic dopants or under surface magnetic deposition \cite{Chen Science Fe, Andrew Nature physics Fe, STM Fe}. Although suppression of photoemission measured density of states at the Dirac point has been reported and interpreted as the magnetic gap \cite{Chen Science Fe, Andrew Nature physics Fe, Ando QPT}, a number of other factors, such as spatial fluctuation of momentum and energy near the Dirac point \cite{Haim Nature physics BiSe} and surface chemical modifications \cite{Andrew Nature physics Fe}, contribute to the observed gap \cite{Haim Nature physics BiSe, Helical metal, SOM}. The photoemission probe previously used to address the gap cannot distinguish or isolate these factors that respect TR symmetry from the TR breaking effect as highlighted in recent STM works \cite{Haim Nature physics BiSe}. In fact, photoemission Dirac point spectral suppression including a gap is also observed even on stoichiometric TI crystals without magnetic dopants or ferromagnetism \cite{SOM}. This is because surface can acquire nontrivial energy gaps due to ad-atom hybridization, surface top layer relaxation, Coulomb interaction from deposited atoms, and other forms of surface chemistry such as in situ oxidation \cite{Andrew Nature physics Fe, Hofmann}. Under such conditions, it was not possible to isolate TR breaking effect from the rest of the extrinsic surface gap phenomena \cite{Andrew Nature physics Fe, Hofmann}. Fundamentally, TR symmetry is inherently connected to the Kramers' degeneracy theorem which states that when TR symmetry is preserved, the electronic states at the TR invariant momenta have to remain doubly spin degenerate. Therefore, the establishment of TR breaking effect fundamentally requires the measurements of electronic groundstate with a \textit{spin} and momentum sensitive probe. Here we utilize spin-resolved angle-resolved photoemission spectroscopy (SR-ARPES) to directly measure the spin configuration and their contrast in systematically magnetically doped, nonmagnetically doped, and ultra-thin TI films \cite{Xue Nature physics QL}. These systematic measurements allow us to \textit{isolate} the TR breaking effect or the magnetic contribution to the surface electronic structure from extrinsic changes in the surface states.

Fig. 1 shows representative SR-ARPES data on Mn-Bi$_2$Se$_3$ MBE films. Our data shows that $P_z$ to be nearly zero at large momentum $k_{//}$ far away from the Dirac point energy ($0<E_B<0.1$ eV in Fig. 1c, d).  When approaching the Dirac point ($0.1$ eV$<E_B<0.3$ eV), an imbalance between spin-resolved intensity in $+\hat{z}$ and $-\hat{z}$ is observed consistently (Fig. 1c) in all samples measured. The imbalance is observed to become more pronounced in data set scans taken by lowering the energy toward the Dirac point. This systematic behavior observed in the data reveals a significant net out-of-plane spin polarization in the vicinity of the ``gapped'' Dirac point or near the bottom of the surface states conduction band, as seen in Fig. 1d. The in-plane spin polarization results ($P_{//}$) are shown in supplementary information \cite{SOM} and found to preserve the helical spin texture (\cite{David Nature tunable}). The observed out-of-plane spin polarization does not reverse its sign in traversing from $-k_{//}$ to $+k_{//}$ and hence cannot be understood as the warping effect in a spin-helical texture \cite{Liang Fu Warping}. 
In the warped helical spin texture case ($P_z$ presents), all three components of the spin vector including $P_z$ are expected to reverse their sign at opposite momenta ($\vec{P}_{x,y,z}(-k)=-\vec{P}_{x,y,z}(k)$), as required by the TR symmetry, which is in sharp contrast to the out-of-plane spin component here in Fig. 1d. In order to directly measure the spin of the surface states at $\bar{\Gamma}$ and obtain a clear picture of the spin texture, we perform spin measurements on Mn-Bi$_2$Se$_3$ film II with spin-resolved energy distribution curves (SR-EDC) mode. The measured out-of-plane spin polarization ($P_z$) is shown in Fig. 1g. We focus on the critical $P_z$ measurements at $\bar{\Gamma}$, $k_{//}=0$ (red curve): the surface electrons at TR invariant $\bar{\Gamma}$ are clearly observed to be spin polarized in out-of-plane direction. The opposite sign of $P_z$ for the upper and lower Dirac band (red curve in Fig. 1g) shows that the Dirac point spin degeneracy is indeed lifted up ($E(k=0,\uparrow){\neq}E(k=0,\downarrow)$). Such observation directly counters the Kramers' degeneracy theorem and therefore breaks the TR symmetry. Next we show $P_z$ measurements along $E_B$ at finite $k_{//}$ (green curves in Fig. 1g) to reveal the detailed configuration of the spin texture. On going to larger $k_{//}$ away from the $\bar{\Gamma}$ momenta, again, the measured $P_z$ is found to decrease gradually to zero. Moreover, the constant energy plane at the Dirac point ($E_B=E_D$) is observed to serve as a mirror plane that reflects all of the out-of-plane spin components between the upper and lower Dirac bands. The spin texture observed here shows a dramatic reorientation as compared to the classical helical spin texture \cite{David Nature tunable} in undoped and nonmagnetic Bi$_2$Se$_3$ system, revealing the breaking of TR symmetry at the Kramers' point ($E(k=0,\uparrow){\neq}E(k=0,\downarrow)$). For systematic comparison we also studied spin texture measurements on analogously nonmagnetically doped TI films (Zn-Bi$_2$Se$_3$). We observe that the out-of-plane spin polarization $P_z$ measurements on the nonmagnetic Zn-Bi$_2$Se$_3$ film show a sharp contrast to the magnetically doped Mn-Bi$_2$Se$_3$ film - near absence of out-of-plane polarization around $\bar{\Gamma}$ within our experimental resolution (Fig. 1j). However, a very small $P_z$ at large $k_{//}$ is expected due to the surface state warping (Fig. 2d). Our in-plane spin measurements (Fig. 2a-c) show that Zn-Bi$_2$Se$_3$ film possesses a helical spin texture \cite{David Nature tunable} consistent with weak warping \cite{Liang Fu Warping} but protected TR symmetry. Based on these systematic studies, our data suggests that Fig. 1 and 2 present clear evidence for TR symmetry breaking in our magnetically doped Mn-Bi$_2$Se$_3$ system. 


We utilize SR-ARPES to measure the spin configuration on the very top (within 5 $\textrm{\AA}$ \cite{SOM}) of the 3QL undoped Bi$_2$Se$_3$ film under quantum coupling as shown in Fig. 2e-h. At large $k_{//}$ far away from $\bar{\Gamma}$ (e.g. -0.10 $\textrm{\AA}^{-1}$ in Fig. 2g), we observe clear spin polarization following left-handed helical configuration with the magnitude of the polarization around $35\sim40$\%. However, when going to smaller  $k_{//}$, the magnitude of spin polarization is observed to be reduced. At the TR invariant $\bar{\Gamma}$ momenta, SR-ARPES measurements (Fig. 2g red curve) show no net spin polarization. The reduction of the spin polarization at small momentums near the gap is physically meaningful, rather than just blurring between the two branches of the upper Dirac cone due to momentum resolution of the instrument. We further prove the momentum resolution by showing that SR-ARPES measurements on 60QL undoped Bi$_2$Se$_3$ film with identical experimental conditions do not show any spin polarization deduction at small momentums (see \cite{SOM}). These observations can be understood by considering the picture that near $\bar{\Gamma}$ the coupling dominates and the two energetically degenerate surface states from top and bottom which possess opposite helicity of the spin texture cancel each other, resulting in strong suppression of spin polarization in the vicinity of a gap, whereas when going to large $k_{//}$ away from $\bar{\Gamma}$, the finite kinetic energy of the surface states (${\propto}vk_{//}$) can lead to the spatial decoupling of the two Dirac cones. 
\textit{Our spin measurements on the ultra-thin Bi$_2$Se$_3$ film shows the quantum tunneling (coupling) effect of ultra-thin TI films in the spin channel}. The observed reconfiguration in ultra-thin Bi$_2$Se$_3$ spin texture, in contrast to the magnetic doping case, does not break TR symmetry, since the spin remains doubly degenerate at the TR invariant momenta $\bar{\Gamma}$. Therefore, our systematic data shows that spin reorientation transition can be realized on the topological surfaces of Bismuth-based TI films via bulk magnetic doping or ultra-thin film quantum tunneling. These contrasting spin groundstates of the surface states under different dopings and conditions are summarized in Fig. 4g-i.

When TR symmetry is broken, a gap is expected to open at the Dirac point that experimentally manifests itself as a spectral weight suppression (SWS) in SI-ARPES measurements. However, as we will show below, the SWS at the Dirac point can be caused not only by TR breaking effect, but also by other factors that are TR invariant. Fig. 3a shows the SI-ARPES measured dispersions of Mn(Zn)-Bi$_2$Se$_3$ film surfaces. Indeed, a SWS is observed with increasing Mn concentrations, whereas the Zn-Bi$_2$Se$_3$ does not show any obvious SWS except at the very large doping concentration (Zn=10\%). The observed SWSs are obvious seen in the raw ARPES and EDC datasets (Fig. 3a-b), independent of the fitting method. In fact, realistic models predict that the Dirac bands will not have regular Voigt profiles near the Dirac point, and there is currently no simple and rigorous way to fit them. Thus we adopt a simpler metric to evaluate the likely amplitude of a gap by defining the energy scale of SWS ($E_{SWS}$) as the energy spacing between the bottom of the upper Dirac band and the Dirac point energy (see Fig. 3b caption and \cite{SOM} for details). The obtained $E_{SWS}$ (Fig. 3c) is useful in consistently evaluating the suppression trend at different doping concentrations. However, the SWS is indeed also observed on Zn=10\%-Bi$_2$Se$_3$ (Fig. 3a), Mn-Bi$_2$Se$_3$ at 300 K (Fig. 3d) where ferromagnetic order already vanishes (see inset of Fig. 4b), and various nonmagnetic stoichiometric TI crystals \cite{SOM}. Therefore, our data suggests that the SWS can arise from both TR breaking (magnetic) and TR invariant (nonmagnetic) origins in any samples (magnetic or nonmagnetic). We further illustrate the physics scenarios that lead to a SWS without breaking TR symmetry in \cite{SOM}. The observation of a SWS (a "gap" in ARPES measurement) at the Dirac point does not necessary lead to the conclusion of TR symmetry breaking even if the samples are magnetically doped. It is the critical observation of the out-of-plane spin polarization near the gap edge (Fig. 1) presented above that for the first time reveals the signature of the TR breaking on the topological surface states of TIs. Although the SWS can arise from both TR breaking (magnetic) and TR invariant (nonmagnetic) origins, the magnetic contribution in a magnetically doped TI film can be isolated using the measured spin texture as demonstrated here. We quantify the magnetic contribution to the SWS by an energy scale $b_z$ which measures the surface magnetic interaction strength (see Fig. 4d caption for $b_z$ definition). Then we define a spin vector out-of-plane polar angle $\theta$ (Fig. 1k). $\theta$ can be directly obtained from the measured spin texture by $\theta=arctan\frac{P_{z}}{P_{//}}$. The polar angle $\theta$ reveals the competition between the out-of-plane TR breaking texture (${\propto}b_z$) and the in-plane helical configuration (${\propto}vk_{//}$). Thus, $b_z$ can be evaluated by $tan\theta=\frac{b_z}{vk_{//}}$ for Mn-Bi$_2$Se$_3$ film II using the SR data in Fig. 1g, as demonstrated in Fig. 4d. Due to the inhomogeneity of the Mn concentration on the film surface (Fig. 4a), $b_z$ obtained from spin texture measurements is a spatial averaged magnetic interaction on the surface. The obtained $b_z$ (21 meV) is much smaller than the corresponding $E_{SWS}$ ($>50$ meV) in Fig. 3c, which again indirectly supports the complex multiple origins of the Dirac point SWS in SI-ARPES. 

We independently measure the magnetic properties of the Mn-Bi$_2$Se$_3$ film surface by the standard methods of X-ray magnetic circular dichroism (XMCD) utilizing the total electron yield (TEY) mode \cite{XMCD, XMCD2}. A clear hysteresis is observed in our XMCD data in Fig. 4b-c. The observed hysteresis loop and spin reorientation observed in SR data together supports a magnetized surface (detailed measurements and discussions are in \cite{SOM}). We also note that the presence of ferromagnetism is revealed in more conventional magnetoresistance and magnetometry measurements, although these cannot unambiguously disentangle bulk from surface effects \cite{Duming}. The observation of out-of-plane spin polarization near the gap edge by SR-ARPES and the XMCD hysteresis indicates that magnetic domains in ferromagnetic thin films are typically large \cite{XMCD, XMCD2}. Magnetic atoms are likely to be inhomogeneously distributed on the surface. However, our data suggests that surface states acquire TR symmetry breaking at least via the proximity effect from the inhomogeneously distributed magnetic Mn atoms on the surface. It is the realization of TR symmetry breaking and associated spin rearrangement on the surface states that is the central requirement for novel device concepts. Theoretically, the TR symmetry breaking can be achieved either by a homogeneous magnetic field or by an inhomogeneous magnetic field as long as their is a net out-of-plane component of the magnetic field. In our case, it is likely that an inhomogeneous Mn atom distribution leads to the TR breaking we observe in the spin texture measurements.

\newpage

\noindent
\textbf{\large{Figure Captions}}

\bigskip

\noindent
\textbf{Figure 1 $|$ Observation of time-reversal symmetry breaking. a-d, }Spin-resolved ARPES measurements on 2.5\% Mn-Bi$_2$Se$_3$ film I using photon energy 20 eV, at T=50K. \textbf{a and b,} Spin-integrated ARPES mapping and momentum distribution curves (MDCs) at incident photon energy 20 eV (used for spin-resolved measurements). The energies selected for spin-resolved measurements are noted. \textbf{c,} Spin-resolved MDC spectra for out-of-plane direction. \textbf{d,} Measured out-of-plane component of the spin polarization. The out-of-plane polar angles ($\theta$) of the spin vectors are noted. \textbf{e-g,} Spin-resolved ARPES measurements on 2.5\% Mn-Bi$_2$Se$_3$ film II using photon energy 9 eV. \textbf{e-f,} Spin-integrated ARPES dispersion and EDCs. The EDCs selected for spin-resolved measurements are highlighted in green and red colors. The EDC at $\bar{\Gamma}$ momenta is in red color. \textbf{g,} Measured out-of-plane spin polarization of the corresponding EDCs. The momentum of each spin-resolved EDC is noted on the top. The out-of-plane polar angles ($\theta$) of the spin vectors are also noted for the upper Dirac cone. \textbf{h-j,} Out-of-plane spin-resolved ARPES measurements on 1.5\% nonmagnetic Zn-Bi$_2$Se$_3$ film with incident photon energy 9 eV. No out-of-plane spin polarization $P_z$ is observed near the TR invariant $\bar{\Gamma}$ momenta. \textbf{k,} Schematic drawing shows the definition of spin polarization vector $\vec{\mathbf{P}}$, out-of-plane component $P_z$ and the out-of-plane polar angle $\theta$.

\bigskip

\bigskip

\bigskip

\bigskip

\noindent
\textbf{Figure 2 $|$ Spin configuration of surface state gap without breaking time-reversal invariance. a-d,} Spin-resolved ARPES measurements on 1.5\% nonmagnetic Zn-Bi$_2$Se$_3$ film with incident photon energy 9 eV. The in-plane spin polarization measurements show the well-established helical spin configuration \cite{David Nature tunable}. \textbf{e-h,} Spin-resolved ARPES measurements on ultra-thin 3QL undoped Bi$_2$Se$_3$ film with incident photon energy 60 eV. Quantum tunneling between the top and bottom surfaces is non-negligible. The two degenerate states with left-hand chirality (LHC) or right-hand chirality (RHC) are originally from top and bottom surfaces. SR-ARPES measures the spin polarization on the very top of the film surface (ARPES typical penetration depth $\sim$5 $\textrm{\AA}$ at photon energy 60 eV) \cite{SOM}. 

\bigskip

\bigskip

\bigskip

\bigskip

\noindent
\textbf{Figure 3 $|$ Observation of strong spectral weight suppression (SWS) in time-reversal broken surface states. a,} High-resolution ARPES measured dispersion of Mn(Zn)-doped Bi$_2$Se$_3$ MBE thin film along $\bar{M}-\bar{\Gamma}-\bar{M}$ direction. The doping level (shown in the top-right corner of each panel) is the nominal Mn(Zn) concentration, which is defined as the ratio of $\frac{Mn(Zn)}{Mn(Zn)+Bi}$ over the entire film crystal. \textbf{b,} The spectral weight suppression (SWS) energy scale $E_{SWS}$ is defined as the energy spacing between the upper Dirac band minimum and the Dirac point energy. The upper Dirac band minimum is obtained by fitting the ARPES measured dispersion by the $\mathbf{k}{\cdot}\mathbf{p}$ \cite{Liang Fu Warping} theory. The fitting procedure of $E_{SWS}$ is illustrated at the lower panels and systematically explained in \cite{SOM}. \textbf{c,} The SWS energy scale $E_{SWS}$ and inverse momentum width $1/{\Delta}k$ are shown as a function of Mn and Zn concentrations at T=20 K. \textbf{d,} Temperature dependence of SWS at the Dirac point of Mn(2.5\%)-Bi$_2$Se$_3$ film. SWS decreases when rising the temperature. However, the suppression survives even up to 300 K where the ferromagnetic order already goes away (see inset of Fig. 4b), which reveals the complex origin of the SWS at the Dirac point.

\bigskip

\bigskip

\bigskip

\bigskip

\newpage
\begin{figure*}
\centering
\includegraphics[width=17cm]{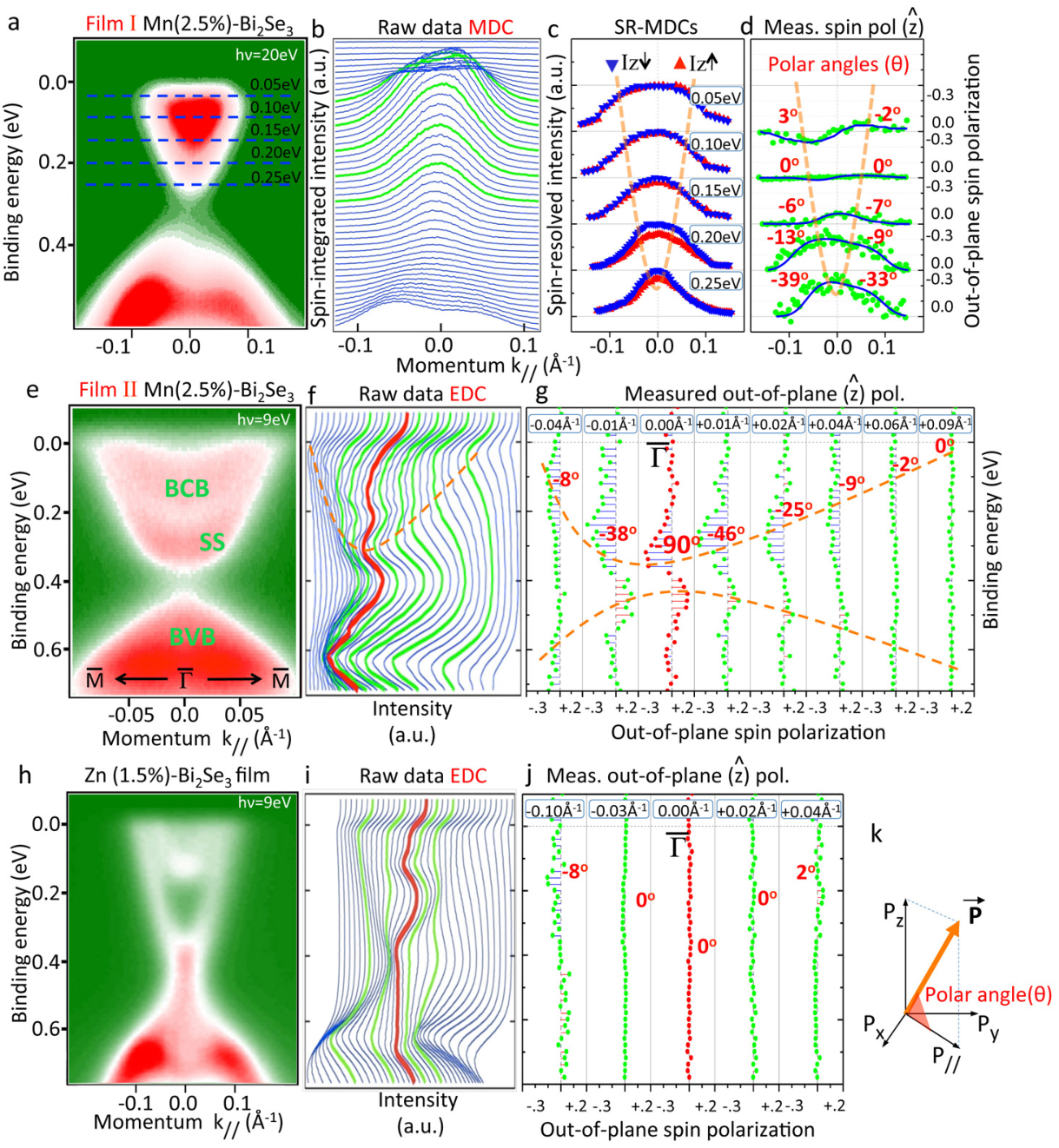}
\end{figure*}

\newpage
\begin{figure}[t]
\centering
\includegraphics[width=17cm]{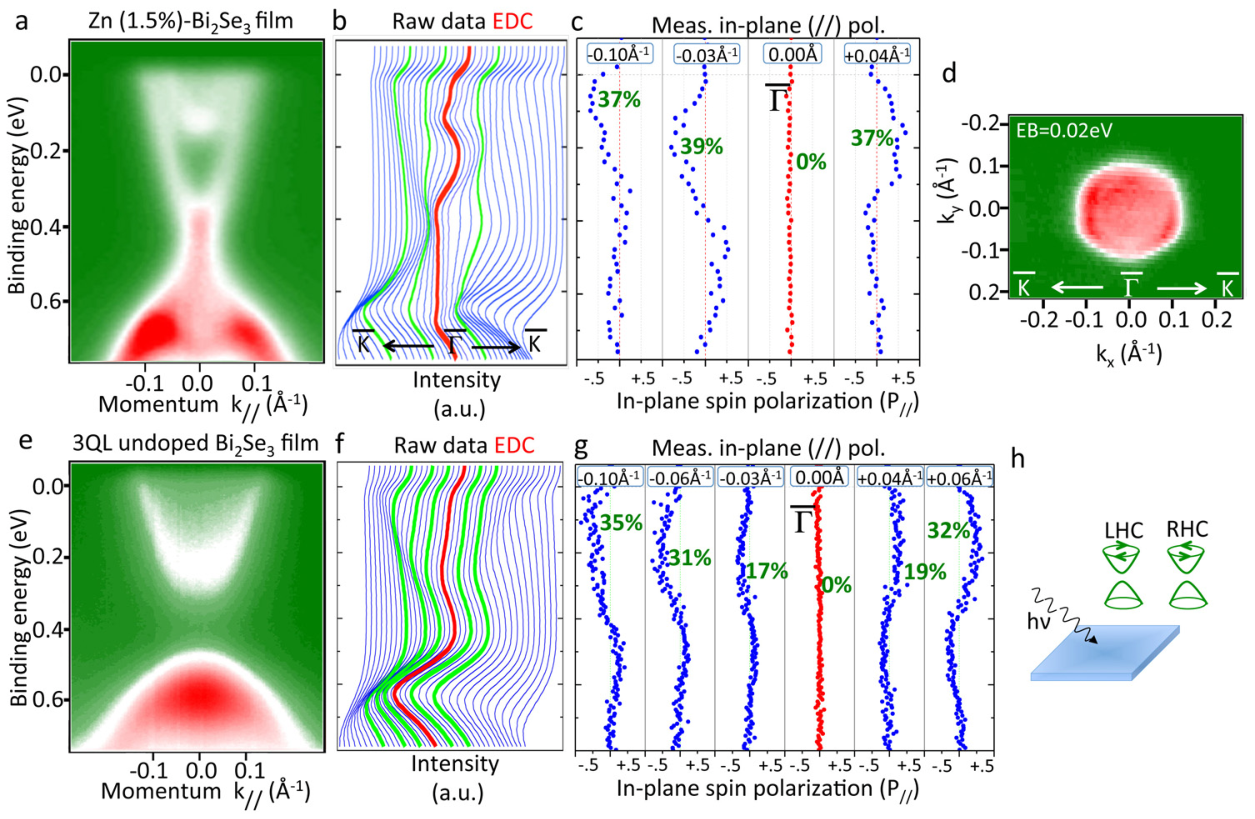}
\end{figure}

\newpage
\begin{figure*}
\includegraphics[width=17cm]{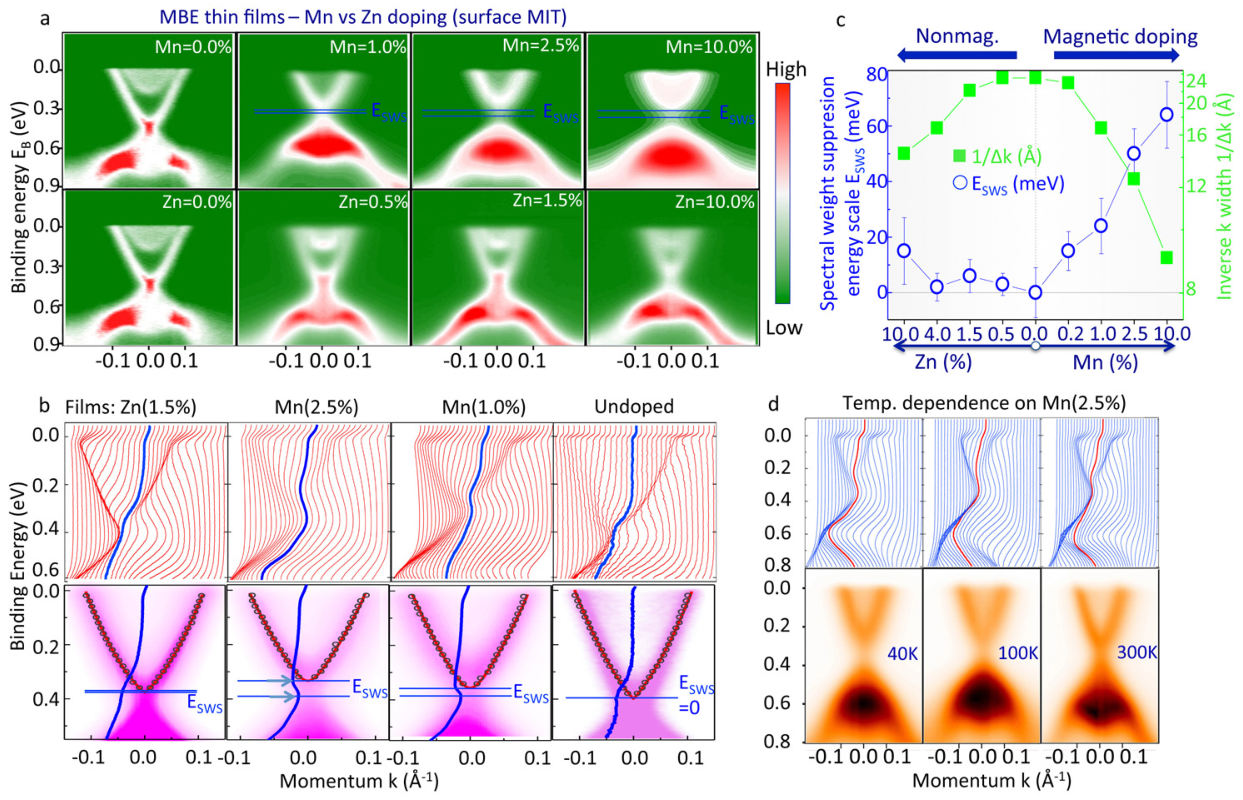}
\end{figure*}


\begin{thebibliography}{21}
\bibitem{SOM} Materials and methods are available as supporting materials.
\end{thebibliography}
\end{document}